\documentclass{article}
\usepackage[T1]{fontenc}
\usepackage{color}

\usepackage{graphicx}
\usepackage{subfigure}
\usepackage{subfigure}

\usepackage{amssymb}
\usepackage{amsthm}
\usepackage{amsmath}

\usepackage{bm}
\usepackage[mathscr]{eucal}
\usepackage{cancel}

\usepackage{wasysym}

\usepackage{tensor}

\def\^0{\:{}^0\!}

\newcommand{\D}{\nabla}
\newcommand{\textfrac}[2]{{\textstyle\frac{#1}{#2}}}

\DeclareMathOperator{\tr}{tr}

\theoremstyle{plain}
\newtheorem{theorem}{Theorem}[section]
\newtheorem{corollary}[theorem]{Corollary}
\newtheorem{lemma}[theorem]{Lemma}

\theoremstyle{remark}

\newtheorem*{remark}{Remark}

\author{\sc
J.\ Mark Heinzle$^{1}$\thanks{Electronic address:
{\tt mark.heinzle@univie.ac.at}} \ and
Patrik Sandin$^{2}$\thanks{Electronic address: {\tt patrik.sandin@kau.se}}\\
$^{1}${\small\em University of Vienna, Faculty of Physics, Gravitational Physics,}\\
{\small\em Boltzmanngasse 5, 1090 Vienna, Austria}\\
$^{2}${\small\em Department of Physics, University of Karlstad,}\\
{\small\em S-651 88 Karlstad, Sweden}}

\title{The initial singularity of ultrastiff perfect fluid spacetimes without symmetries}
\begin{document}
\maketitle

\begin{abstract}
We consider the Einstein equations coupled to an ultrastiff
perfect fluid and prove the existence of a family of solutions
with an initial singularity whose structure is that of explicit
isotropic models. This family of solutions is `generic' in the
sense that it depends on as many free functions as a general
solution, i.e., without imposing any symmetry assumptions, of
the Einstein-Euler equations. The method we use is a that of a
Fuchsian reduction.
\end{abstract}

\section{Introduction}

In general relativity, spacetime singularities can occur under
very general conditions, see~\cite{hawkingpenrose}. However,
mathematical theorems detailing the structure of these
singularities usually assume a great deal of symmetry; either
spatial homogeneity is assumed, or as in the Gowdy and $T^2$
cases, an Abelian two-parameter isometry group acting on
spatial sections. A potentially powerful method to obtain
theorems concerning non-homogeneous models is the method of
Fuchsian reduction~\cite{Kichenassamy}, which is a rigorous attempt to `expand' solutions in the neighborhood of a singularity.
A particular way (but by no means the only one~\cite{Kichenassamy}) to obtain the leading order behavior of solutions
toward a singularity is
to formulate an `asymptotic system' of equations
that is
derived from the Einstein equations by dropping
terms, e.g., spatial derivatives,
which are believed to be unimportant for the leading asymptotic
dynamics. If this `asymptotic system' truly captures the
leading divergent behavior of solutions of the Einstein
equations toward a singularity, then the remaining part of each solution obeys
equations that are of so-called Fuchsian type. In this case, it
is possible to prove the existence of a general class of
solutions of the Einstein equations whose asymptotic behavior
is governed by the behavior of solutions of the asymptotic
system. This method of Fuchsian reduction was applied, e.g.,
in~\cite{Andersson&Rendall} to the initial singularity of
cosmological models with a stiff fluid or massless scalar field
in $3+1$ dimensional spacetimes without symmetries, and to the
Einstein-dilaton-$p$-form system in higher
dimensions~\cite{Damour/Henneaux/Rendall/Weaver:2002}.

The structure of spacetime singularities is particularly
interesting in the context of the Weyl curvature hypothesis,
where the focus is on isotropic
singularities~\cite{Anguine&Tod}. Spacetimes with such
singularities are not generic among solutions of the
Einstein-matter equations, i.e., an arbitrary matter/curvature
distribution at late times does not develop an isotropic
initial singularity when traced back in time. However, this is
expected to only hold for matter sources where the ratio of
pressure to energy density is less than or equal to one; if
this ratio is greater than one, isotropic singularities are
expected to be generic. In the case of a perfect fluid matter
source, such a matter model is called an ultrastiff perfect
fluid.

The interest in ultrastiff fluids has come from attempts to
obtain non-singular descriptions of spacetimes representing
collapsing universes bouncing outward before a singularity
occurs. This in the case in the ecpyrotic scenario or the
cyclic brane world models~\cite{Khouryetal}, where the
ultrastiff fluid is an effective description of a scalar field
with negative potential energy during the contraction phase.
The scalar field is in turn a representation of the size of a
compactified dimension and the potential a representation of
the force between two different brane worlds. These models have
led to an interest in ultrastiff cosmologies in
general~\cite{Barrow&Yamamoto} and especially in their
structure in the final stages of collapse, or equivalently,
initial stages of expansion in an expanding universe. The
structure of the initial singularity in cosmological models
with an ultrastiff perfect fluid has been studied under the
assumption that the singularity is `local'~\cite{Ericksonetal,
heietal09}, which means that the time evolution at adjacent
spatial points decouples. Under this a priori assumption of
locality it has been demonstrated that the past asymptotic
structure is described by an isotropic
solution~\cite{Ericksonetal, Coley&Lim} of the
Friedmann-Robertson-Walker type, i.e., that anisotropies become
small compared to the overall expansion of the universe close
to the singularity. The situation is similar for cosmological
models with several non-interacting perfect fluids, of which at
least one possesses an ultrastiff equation of
state~\cite{sanugg10}.

In this paper we investigate the past asymptotic dynamics of
cosmological spacetimes with ultrastiff perfect fluids. We do
not impose any assumptions on the symmetry of the spacetimes
and we do not make any a priori assumptions on the past
asymptotic behavior (like asymptotic locality). We prove the
existence of a family of solutions that converge to isotropic
models toward the initial singularity; the past asymptotic
behavior of these solution is characterized in detail, see
Theorem~\ref{thm2}. This family of solutions we construct is
`generic' in the sense that it depends on as many free
functions as a general solution of the initial value problem
connected with the Einstein-Euler equations.

The paper is organized as follows. In section~\ref{sec:EEeqs}
we give the Einstein-Euler equations in a Gaussian coordinate
system; in section~\ref{sec:asymptoticsystem} we describe the
asymptotic system and its role in the Fuchsian reduction.
Section~\ref{sec:w=3} is devoted to the special case of a
perfect fluid with equation of state $p=3\mu$, for which the
asymptotic system admits explicit solutions. We use these
solutions to set up a Fuchsian system of equations; this is
done in analogy with the analysis of the stiff fluid
case~\cite{Andersson&Rendall}. In section~\ref{sec:generalw} we
give the treatment of the general ultrastiff fluid case, which
is less explicit but follows the same principles. The main result is then stated in section~\ref{sec:mainresult}, while section~\ref{sec:reducedsystem} gives a derivation of the reduced equations that are used to prove the main result in section~\ref{sec:Fuchsiananalysisgenw}.

\section{The Einstein-Euler equations}\label{sec:EEeqs}

Let ${}^4\!M$ be a four-dimensional Lorentzian manifold with
metric ${}^4\!g_{\alpha\beta}$ (with $\alpha, \beta =
0,1,2,3$). We consider the Einstein-Euler equations for
self-gravitating perfect fluids, i.e.,
\begin{subequations}
\begin{align}
\label{einstein}
& {}^4\!R_{\alpha\beta} -\textfrac{1}{2} \,{}^4\!R \:\,{}^4\!g_{\alpha\beta}
= T_{\alpha\beta} \equiv \mu u_\alpha u_\beta +
p \big( {}^4\!g_{\alpha\beta} + u_\alpha u_\beta \big)\:,\\[0.5ex]
\label{bianchi}
& \nabla_\alpha T^{\alpha\beta} = 0 \:,
\end{align}
\end{subequations}
where we assume a linear equation of state $p = w \mu$,
$w=\mathrm{const}$, relating the energy density $\mu$ (as
measured in the fluid's rest frame) and the pressure $p$ of the
fluid; $u^\alpha$ is the four-velocity of the fluid. We use
geometrized units, i.e., $c = 1$ and $8\pi G = 1$,  where $c$
is the speed of light and $G$ the gravitational constant.

We consider spacetimes that are diffeomorphic to $\mathbb{R}
\times M$, where $M$ is three-dimensional, and metrics of the
form
\begin{equation}\label{metric}
-d t^2 + g_{a b} \,\omega^a \omega^b \:,
\end{equation}
where $g_{a b} = g_{a b}(t)$ denotes a one-parameter family of
Riemannian metrics, which are naturally identified with the
metrics induced on $t = \mathrm{const}$ hypersurfaces; the
coframe $\{\omega^1, \omega^2, \omega^3\}$ on $M$ is arbitrary.
The range of Latin indices $a,b,\ldots$ is $1,2,3$.

The Einstein-Euler equations become a first order system of
equations with constraints when the extrinsic curvature $k_{a
b} = -\textfrac{1}{2}\partial_t g_{a b}$ of the
$t=\mathrm{const}$ hypersurfaces is used ($3+1$ split). For our
purposes it is preferable to split $k_{a b}$ into its trace
(`mean curvature') and (the negative of) its traceless part,
i.e.,
\begin{equation}
k_{a b} = -\sigma_{a b} + \textfrac{1}{3}(\tr k)\, g_{a b}\:.
\end{equation}
The tensor $\sigma_{a b}$ coincides with the rate of shear
tensor of the (geodesic) congruence orthogonal to the
$t=\mathrm{const}$ hypersurfaces. Then the evolution equations
are
\begin{subequations}\label{evol}
\begin{align}
\partial_t g_{a b} & = 2 g_{a c} \big( \tensor{\sigma}{^c_b} -
\textfrac{1}{3}(\tr k)\, \tensor{\delta}{^c_b}\big)\,,\\[0.8ex]
\label{evolsig}
\partial_t \tensor{\sigma}{^a_b} & = (\tr k) \tensor{\sigma}{^a_b} -
\big( \tensor{R}{^a_b} - \textfrac{1}{3} R \tensor{\delta}{^a_b}\big) +
\big( \tensor{S}{^a_b} - \textfrac{1}{3} (\tr S) \tensor{\delta}{^a_b}
\big)\,, \\[0.8ex]\partial_t (\tr k) & = R + (\tr k)^2 +
\textfrac{1}{2}(\tr S)-\textfrac{3}{2}\rho\,,
\end{align}
\end{subequations}
and the constraint equations read
\begin{subequations}\label{constr}
\begin{align}
& R - \tensor{\sigma}{^a_b}\tensor{\sigma}{^b_a} + \textfrac{2}{3}
(\tr k)^2= 2\rho,\\[0.8ex]
&-\nabla_a \tensor{\sigma}{^a_b} -\textfrac{2}{3} \nabla_b(\tr k) =  j_b \:,
\end{align}
\end{subequations}
where
\begin{equation}\label{rhojs}
\rho =\mu \big( 1 +(1+w) u^a u_a \big)\,,\quad
j_b =\mu (1 + w) (1+u_a u^a)^{1/2} \,u_b\,,\quad
S_{a b} =\mu \big((1+w)u_a u_b+w g_{ab} \big)\:.
\end{equation}
The quantity $\rho$ is the energy density, $j_a$ the current
measured by the observer associated with the geodesic
congruence. Equations~\eqref{evol} and~\eqref{constr}, together
with~\eqref{rhojs}, correspond to~\eqref{einstein}. The Euler
equations~\eqref{bianchi} read
\begin{subequations}\label{euler}
\begin{align}
\nonumber
\partial_t\mu -(1+w)(\tr k) \mu & =
-(1+w)
\Big[ u^2 \partial_t\mu + 2\mu u^a \partial_t u_a
- \mu \sigma_{a b} u^a u^b - \textfrac{2}{3} \mu (\tr k) u^2 \\
\label{eulermu}
& \qquad
+ \sqrt{1 + u^2} \:u^a \D_a \mu
+ \mu (1 + u^2)^{-1/2} u_a u^b \D_b u^a
+ \mu \sqrt{1 + u^2}\, \D_a u^a \Big]\,,
\end{align}
\begin{align}
\nonumber
\partial_t u_a  + w\, (\tr k) \,u_a + \textfrac{w}{1+w} \mu^{-1} \D_a\mu  & =
-\mu^{-1} u_a \big[ \partial_t\mu - (1+w)\mu(\tr k)\big]
- u_a(1 + u^2)^{-1} \big[ u^b\partial_t u_b - \sigma_{b c}u^b u^c \big] \\[0.8ex]
\nonumber
& \qquad
+ \textfrac{1}{3} u_a(1 + u^2)^{-1} (\tr k) u^2
- [(1 + u^2)^{-1/2}-1]\textfrac{w}{1+w} \mu^{-1} \D_a\mu \\[0.8ex]
\label{euleru}
& \qquad
- (1 + u^2)^{-1/2} \big[ \mu^{-1} u_a u^b \D_b\mu + u_a\D_b u^b + u^b \D_b
u_a\big]\;,
\end{align}
\end{subequations}
where we use the abbreviation $u^2 = u_a u^a$.

\section{The asymptotic system}\label{sec:asymptoticsystem}

Alongside the Einstein-Euler equations we consider another
system of equations for the same set of variables, which we
denote by $\big(\^0g_{a b}, \tr \^0k, \^0\tensor{\sigma}{^a_b},
\^0\mu,\^0u_a\big)$ in the present context. This
\textit{asymptotic system} consists of evolution equations for
$\big(\^0g_{a b}, \tr \^0k, \^0\tensor{\sigma}{^a_b}\big)$,
\begin{subequations}\label{evoleqs}
\begin{align}
\label{evolg0}
\partial_t \^0g_{a b} & =
2 \^0g_{a c} \big( \^0\tensor{\sigma}{^c_b} - \textfrac{1}{3} (\tr \^0k)
\tensor{\delta}{^c_b} \big) \,,\\[0.5ex]
\label{evolsig0}
\partial_t \^0\tensor{\sigma}{^a_b} &= (\tr  \^0k) \^0 \tensor{\sigma}
{^a_b}\:, \\[0.5ex]
\label{evolk0}
\partial_t (\tr  \^0k) &= (\tr  \^0k)^2 + \textfrac{3}{2}(w-1)\^0\mu\,,
\end{align}
\end{subequations}
and evolution equations for the matter variables $(\^0\mu,\^0u_a)$,
\begin{subequations}\label{mattereqs}
\begin{align}
\label{evolmu0}
\partial_t\^0\mu - (1+w)(\tr  \^0k) \^0\mu &= 0\,,\\[0.5ex]
\label{evolu0}
\partial_t\^0u_a + w \,(\tr \^0k) \^0u_a + \textfrac{w}{1+w}\,\^0\mu^{-1}\,\^0\nabla_a\^0\mu &= 0 \:.
\end{align}
\end{subequations}
In addition, we consider the constraint equations
\begin{subequations}\label{constraint equations}
\begin{align}
\label{hamcons}
-\^0\sigma^a{}_b\^0\sigma^b{}_a + \textfrac{2}{3}(\tr  \^0k)^2 & =
2\^0\rho,,\\[0.5ex]
\label{momcons}
-\^0\nabla_a \^0\tensor{\sigma}{^a_b}- \textfrac{2}{3} \^0\nabla_b
(\tr \^0k)&= \^0j_b \:,
\end{align}
\end{subequations}
where $\^0\rho \equiv \^0\mu$ and $\^0j_a \equiv (1+w) \^0\mu
\^0u_a$. The symbol $\^0\nabla_a$ denotes the covariant
derivative associated with the metric $\^0g_{a b}$.

Note that the evolution equations~\eqref{evoleqs}
and~\eqref{mattereqs} preserve the
constraints~\eqref{constraint equations}. A computation shows
that
\begin{align*}
\partial_t \big( {-\^0\sigma^a{}_b\^0\sigma^b{}_a} + \textfrac{2}{3}(\tr
\^0k)^2 - 2\^0\rho \big)
& = 2 (\tr \^0k) \big( {-\^0\sigma^a{}_b\^0\sigma^b{}_a} + \textfrac{2}{3}
(\tr  \^0k)^2 - 2\^0\rho \big) \,,\\[0.8ex]
\partial_t \big({-\^0\nabla_a \^0\tensor{\sigma}{^a_b}}- \textfrac{2}{3}
\nabla_b (\tr \^0k) - \^0j_b \big)
& = (\tr\^0k) \big({-\^0\nabla_a \^0\tensor{\sigma}{^a_b}}- \textfrac{2}{3}
\^0\nabla_b (\tr \^0k) - \^0j_b \big) \\
& \qquad \qquad\qquad -\textfrac{1}{2} \^0\nabla_b \big( {-\^0\sigma^a{}_c}\^0
\sigma^c{}_a + \textfrac{2}{3}(\tr  \^0k)^2 - 2\^0\rho \big) \,,
\end{align*}
hence if the constraints are satisfied initially, they are
satisfied for all times.

The asymptotic system~\eqref{evoleqs}--\eqref{constraint
equations} is a system of partial differential equations,
where, however, the spatial derivatives merely enter in a
passive manner, through the decoupled equations~\eqref{evolu0}
and~\eqref{momcons}; hence in
spirit,~\eqref{evoleqs}--\eqref{constraint equations} are ODEs.
It is clear that the asymptotic system is obtained from the
Einstein-Euler system by dropping (a large number of) terms.
The derivation of the asymptotic system from the Einstein-Euler
system is intimately connected with considerations involving
the concepts of asymptotic silence and locality; we refer
to~\cite{heietal09} and references therein. The aim of these
considerations is to find a simple system of equations that
reproduces the dynamics of the full Einstein-matter equations
in the asymptotic limit toward a `generic spacelike
singularity'. In other words, a simple (`asymptotic') system is
sought that governs the asymptotic behavior toward a spacelike
singularity of (typical) solutions of the Einstein-matter
equation in the sense that each (typical) solution of the
Einstein-matter equation is approximated, to a certain order,
by a solution of the asymptotic system.

Whether the asymptotic system given
in~\eqref{evoleqs}--\eqref{constraint equations} indeed governs
the asymptotic behavior of solutions of the full Einstein-Euler
equations, is a non-trivial question. The considerations
of~\cite{heietal09} strongly indicate that a necessary
condition is that the equation of state parameter satisfies $w
\geq 1$. (In the case $w < 1$, the conjectured asymptotic
system is considerably more complicated, because it must
capture the conjectured oscillatory behavior, `Mixmaster
behavior', of solutions toward the singularity.) The case of a
stiff fluid, i.e. $w = 1$, has been treated
in~\cite{Andersson&Rendall}: The asymptotic system given
in~\cite{Andersson&Rendall} is shown to govern the asymptotic
behavior of solutions of the Einstein-Euler equations with a
stiff fluid. (Note, however, the comment in the symmetry
subsection of section~\ref{sec:generalw}.)

In this paper we consider ultrastiff fluids, which are
characterized by $w > 1$. We prove that the asymptotic behavior
of solutions (more specifically, of a full set of solutions in
the sense of function counting) of the Einstein-Euler equations
with an ultrastiff stiff fluid is governed by the asymptotic
system~\eqref{evoleqs}--\eqref{constraint equations}; in
addition, we give detailed information about the order of the
approximation.

\subsection{A particular(ly simple) case of ultrastiff fluids:
$\bm{w=3}$}\label{sec:w=3}

We begin by solving the asymptotic system with $w = 3$. The
reason for this choice is one of convenience, since in the special
case $w = 3$, the solutions of the asymptotic system take a
simple and explicit form (which enables us to stress the
analogies of the present analysis with the analysis
of~\cite{Andersson&Rendall}).
The general (but less explicit) case $w >1$ is treated in subsection~\ref{sec:generalw}.

We are interested in solutions
with $\^0\mu > 0$ and $\tr \^0k < 0$, i.e., `expanding' models
with positive energy density. (Note that $\^0\mu$ remains
positive if it is positive initially, see~\eqref{evolmu0};
likewise, if $\tr \^0 k$ is negative at $t = t_0$, it remains
negative for all $t < t_0$ because $\partial_t(\tr\^0k) > 0$,
see~\eqref{evolk0}.) The basic observation is that
\begin{equation}\label{pm}
\partial_t (\sqrt{3\^0\mu} \pm \tr \^0k) = \pm (\sqrt{3\^0\mu} \pm \tr\^0k)^2.
\end{equation}
The solutions of~\eqref{pm} with the minus sign are
\begin{equation}\label{m}
\sqrt{3\^0\mu} - \tr \^0k = \frac{1}{t + \psi(x)}\,,
\end{equation}
with $t > -\psi(x)$, where $\psi = \psi(x)$ is an arbitrary
function of the spatial coordinates which we collectively
denote by $x$. (The trivial solution is excluded by our
assumptions.) The `initial singularity' is signalled by the
divergence of~\eqref{m} at $t = -\psi$. We wish this initial
singularity to occur at $t = 0$. To simplify further
computations we thus choose our initial hypersurface to be a
surface where $\psi = \mathrm{const}$. This is not a
restriction of the generality of the result since it is always
possible to find such a surface between any given initial
hypersurface and the singularity. The constant may be set to
zero by a shift in the time coordinate.

The solutions of~\eqref{pm} with the plus sign are
\begin{equation}\label{p}
\sqrt{3\^0\mu} + \tr \^0k = -\frac{1}{t + \phi(x)} \quad\text{ and }\quad \sqrt{3\^0\mu} + \tr \^0k \equiv 0 \:,
\end{equation}
where $\phi = \phi(x) > 0$ is an arbitrary function of the
spatial coordinates. (The function $\phi$ is positive, since $2
\sqrt{3\^0\mu} = t^{-1} - (t + \phi)^{-1} > 0$.) The trivial
solution in~\eqref{p} can be thought of as arising as a special
case by taking the limit $\phi\rightarrow \infty$. We conclude
that
\begin{subequations}\label{asybeh}
\begin{align}
&\tr\^0k =-\frac{1}{2}\left(\frac{1}{t} + \frac{1}{t + \phi}\right),
\qquad
\^0\mu = \frac{1}{12}\left(\frac{1}{t} - \frac{1}{t + \phi}\right)^2\,.
\intertext{Integration of the remaining evolution equations
of~\eqref{evoleqs} and~\eqref{mattereqs} then yields}
&\label{asybehsig} \^0\tensor{\sigma}{^a_b} =
\frac{\^0\tensor{\mathring{\sigma}}{^a_b}}{\sqrt{t(1+\phi^{-1}t)}} \:,\\
&\label{asybehu} \^0u_a = \^0\mathring{u}_a\,\big(t\,(1+\phi^{-1}t)
\big)^{3/2} - 3 \phi^{-2}(\nabla_a\phi)\,t^2 \,\big(1 + \textfrac{2}{3}
\phi^{-1} t\big),\\[0.5ex]
&\label{asybehg} \^0g_{ab} = \^0\mathring{g}_{ac} \big(t\,(1+\phi^{-1}t)
\big)^{1/3}\Big( \sqrt{\phi^{-1} t} + \sqrt{ 1 + \phi^{-1} t}
\Big)^{4 \sqrt{\phi} \^0\tensor{\mathring{\sigma}}{^c_b}} \:,\\[1ex]
&\label{asybehginv}\^0g^{ab} = \big(t\,(1+\phi^{-1}t)\big)^{-1/3}
\Big( \sqrt{\phi^{-1} t} + \sqrt{ 1 + \phi^{-1} t} \Big)^{-4 \sqrt{\phi}
\^0\tensor{\mathring{\sigma}}{^a_c}} \, \^0\mathring{g}^{cb}\:.
\end{align}
\end{subequations}
The equations for the metric and its inverse use the matrix
exponential $z^\mathbf{A} := e^{(\log z)\mathbf{A}}$, which is
well defined for any positive scalar $z$ and square matrix
$\mathbf{A}$. Note that $(\sqrt{\phi^{-1} t} + \sqrt{ 1 +
\phi^{-1} t})^{\sqrt{\phi}} \rightarrow e^{\sqrt{t}}$ in the
limit $\phi\rightarrow \infty$, hence $\^0g_{ab} =
\^0\mathring{g}_{ac} t^{1/3} \exp(4\sqrt{t}
\^0\tensor{\mathring{\sigma}}{^c_b})$ in that special case.

The contraints~\eqref{constraint equations} give restrictions
on the free functions of the spatial variables: $\phi$,
$\^0\tensor{\mathring{\sigma}}{^a_b}$, $\^0\mathring{g}_{ab}$,
$\^0\mathring{u}_a$. We have
$\^0\tensor{\mathring{\sigma}}{^a_b}\^0\tensor{\mathring{\sigma}}{^b_a}
= 2\phi/3$ from~\eqref{hamcons}. The constraint~\eqref{momcons}
can be used to express, e.g., $\^0\mathring{u}_a$ in terms of
$\^0\tensor{\mathring{\sigma}}{^a_b}$, $\^0\mathring{g}_{ab}$,
and $\phi$. For the Fuchsian analysis of Section~\ref{sec:Fuchsiananalysisgenw} we will assume that the free functions are analytic in the spatial variables $x$.

\subsection{Ultrastiff fluids with general $\bm{w > 1}$}
\label{sec:generalw}

In the case of a general ultrastiff fluid with an arbitrary
value of $w > 1$ the solutions of the asymptotic system, which are given by~\eqref{asybeh} in the case $w = 3$, cannot be given in
explicit form.
To derive the solutions of
the asymptotic system with $w \neq 3$ we resort to power
series; the ideal framework is the Hubble-normalized dynamical
systems approach.

Consider the asymptotic
system~\eqref{evoleqs}--\eqref{mattereqs}. The
equations~\eqref{evolg0} for $\^0g_{a b}$ and~\eqref{evolu0}
for $\^0u_a$, together with~\eqref{momcons}, decouple from this
system, i.e., the asymptotic system reduces to equations for
$\^0\tensor{\sigma}{^a_b}$, $\tr \^0k$, and $\^0\mu$, which are
subject to the constraint~\eqref{hamcons}.
We write these equations in the standard Hubble-normalized
variables, i.e., we introduce
\begin{equation}
H = -\textfrac{1}{3} \tr \^0k\:\quad\text{and}\quad\:
\tensor{\Sigma}{^a_b} = \frac{\tensor{\^0\sigma}{^a_b}}{H}\,,\:\:
\Omega = \frac{\^0\mu}{3 H^2} \:,
\end{equation}
and an adapted time variable $\partial_\tau = H^{-1}
\partial_t$. In these variables we obtain a decoupled evolution
equation for $H$,
\begin{subequations}\label{Hubbleevol}
\begin{align}
\label{evolH}
\partial_\tau H &= -\textfrac{3}{2}\,H \big(  w + 1 - (w - 1)
\Sigma^2\big)\;,
\intertext{and an evolution equation for $\tensor{\Sigma}{^a_b}$,}
\label{evolSig}
\partial_\tau \tensor{\Sigma}{^a_b} & = \,\textfrac{3}{2} (w - 1)
\tensor{\Sigma}{^a_b} (1 - \Sigma^2)\:,
\end{align}
\end{subequations}
where $\Sigma^2 = \textfrac{1}{6}
\tensor{\Sigma}{^a_b}\tensor{\Sigma}{^b_a}$. The constraint
simplifies to $\Omega = 1 - \Sigma^2$ (which makes the
evolution equation for $\Omega$ redundant).

Equation~\eqref{evolSig} can be solved straightforwardly; we
obtain
\begin{subequations}
\begin{equation}
\frac{\tensor{\Sigma}{^a_b}}{\sqrt{1 - \Sigma^2}} =
\frac{\tensor{\mathring{\Sigma}}{^a_b}}{\sqrt{1 - \mathring{\Sigma}^2}}\:\,
e^{\textfrac{3}{2}(w - 1) \tau}\:,
\end{equation}
where $\tensor{\mathring{\Sigma}}{^a_b}$ are functions of the
spatial variables. Furthermore,~\eqref{Hubbleevol} yields
\begin{equation}
H \sqrt{1 - \Sigma^2} = \mathring{H}\sqrt{1 - \mathring{\Sigma}^2} \:\,
e^{-\textfrac{3}{2}(w + 1) \tau} \:,
\end{equation}
\end{subequations}
where $\mathring{H}$ depends on the spatial variables.
Consequently,
\begin{subequations}\label{intau}
\begin{align}
\^0\tensor{\sigma}{^a_b} &= \mathring{H} \tensor{\mathring{\Sigma}}{^a_b}
e^{-3 \tau} \:, \\
\label{trkintau}
\tr\^0 k & = -3 H = -3 \mathring{H}\,\sqrt{1 - \mathring{\Sigma}^2} \:
e^{-\textfrac{3}{2} (1 +w) \tau}\:
\Big(1 + \frac{\mathring{\Sigma}^2}{1 - \mathring{\Sigma}^2}\:
e^{3 (w -1) \tau}\Big)^{1/2}\,,
\end{align}
\end{subequations}
and $\^0\mu = 3 \mathring{H}^2 (1 - \mathring{\Sigma}^2)\,
e^{-3(w + 1) \tau}$. Accordingly, we obtain explicit expression
for the solutions of the asymptotic system in the time variable
$\tau$.

In order to express~\eqref{intau} in terms of cosmological time
$t$ we must integrate the relation $d t = H^{-1} d \tau$ where
$H$ is given by~\eqref{trkintau}. For large negative $\tau$ we
can expand the square root in a power series in
$e^{3(w-1)\tau}$, and integrate it term by term.
Requiring synchronous cosmological time, where $t = 0$
corresponds to the initial singularity, we obtain
\begin{align*}
t & = \frac{2 \:e^{\textfrac{3}{2} (1 +w) \tau} }{3 (1+w) \mathring{H}
\sqrt{1 - \mathring{\Sigma}^2}}\:\, \times \\
& \qquad\times \left( 1 - \frac{1+w}{2(3 w -1)} \frac{\mathring{\Sigma}^2}
{1 - \mathring{\Sigma}^2} \, e^{3 (w-1) \tau}
+ \frac{3 (1+w)}{8(5 w -3)} \Big(\frac{\mathring{\Sigma}^2}
{1 - \mathring{\Sigma}^2}\Big)^2 \, e^{6 (w-1) \tau}
+ O\big(e^{9 (w-1)\tau} \big) \right)
\end{align*}
as $\tau\rightarrow -\infty$. Inverting this relation results
in
\begin{align*}
\nonumber
e^{\textfrac{3}{2} (1 +w) \tau} & = \textfrac{3}{2} (1+w) \mathring{H}
\sqrt{1 - \mathring{\Sigma}^2} \:\,t\:\,\times \\
& \qquad\times\left( 1 + \frac{1+w}{2}
\Big(\frac{3(1+w)}{2}\Big)^{2(w-1)/(1+w)}
\frac{\mathring{\Sigma}^2  \mathring{H}^{2(w-1)/(1+w)}}{(3 w- 1)
(1 - \mathring{\Sigma}^2)^{2/(1+w)}}\:t^{\textfrac{2(w-1)}{1+w}}
+ O\big( t^{\textfrac{4(w-1)}{1+w}}\big)
\right)
\end{align*}
in the limit $t\rightarrow 0$. A calculation then shows that
\begin{subequations}\label{asybehgenw}
\begin{align}
\nonumber
\^0\tensor{\sigma}{^a_b} &=  \mathring{H} \tensor{\mathring{\Sigma}}{^a_b}
\Big(\textfrac{3}{2} (1+w) \mathring{H}\Big)^{-2/(1+w)}
(1 - \mathring{\Sigma}^2)^{-1/(1+w)} \:\,t^{-\textfrac{2}{1+w}}
\: \Big( 1 + O\big( t^{\textfrac{2(w-1)}{1+w}}\big) \Big)  \\
& \equiv \^0\tensor{\mathring{\sigma}}{^a_b} \: t^{-\textfrac{2}{1+w}}
\: \Big( 1 + O\big( t^{\textfrac{2(w-1)}{1+w}}\big) \Big) \:,
\label{asybehgenwsig}\\[1ex]
\nonumber
\tr\^0 k  & = -3 H = -\frac{2}{1+w}\:t^{-1}
\left(1 +
\frac{w-1}{3 w -1} \Big(\frac{3(1+w)}{2}\Big)^{\textfrac{2 (w-1)}{1+w}} \,
\frac{\mathring{\Sigma}^2 \mathring{H}^{2(w-1)/(1+w)}}
{(1-\mathring{\Sigma}^2)^{2/(1+w)}}
 \: t^{\textfrac{2(w-1)}{1+w}} + O\big( t^{\textfrac{4(w-1)}{1+w}}\big)
\right) \\
& \equiv -\frac{2}{1+w}\:t^{-1}
\left(1 + \phi^{-1} t^{\textfrac{2(w-1)}{1+w}} + O\big( t^{\textfrac{4(w-1)}
{1+w}}\big) \right) \:,\label{asybehgenwtrk}
\end{align}
where we have introduced the spatial functions
$\^0\tensor{\mathring{\sigma}}{^a_b}$ and $\phi$. In
addition,~\eqref{evolg0} and~\eqref{mattereqs} yield
\begin{align}
\^0g_{a b} & = \^0\mathring{g}_{a c} \:t^{\textfrac{4}{3(1+w)}} \,
\left( \tensor{\delta}{^c_b}
+ 2\, \frac{w+1}{w-1}\, \^0\tensor{\mathring{\sigma}}{^c_b} \,
t^{\textfrac{w-1}{w+1}} + O\big( t^{\textfrac{2(w-1)}{1+w}}\big) \right)
\:,\label{asybehgenwg}\\[1ex]
\^0\mu & = \frac{4}{3} \frac{1}{(1 + w)^2} \: t^{-2} \,
\left( 1 - \frac{w+1}{w-1} \,\phi^{-1} t^{\textfrac{2(w-1)}{1+w}} +
O\big( t^{\textfrac{4(w-1)}{1+w}}\big) \right)\:,\label{asybehgenwmu}
\\[1ex]
\^0u_a & = \^0\mathring{u}_a \,t^{\textfrac{2 w}{1+w}}
- \frac{w (w+1)}{(w-1)^2}\phi^{-2} \nabla_a \phi \:\, t^{\textfrac{2 w}{1+w}}
t^{\textfrac{w-1}{1+w}} + t^{\textfrac{2 w}{1+w}}
O\big( t^{\textfrac{2(w-1)}{1+w}}\big) \:, \label{asybehgenwu}
\intertext{and requiring that $\^0g^{ab}$ is the inverse of $\^0g_{ab}$
gives}
\^0g^{a b} & = \^0\mathring{g}^{a c} \:t^{\textfrac{4}{3(1+w)}} \,
\left( \tensor{\delta}{^b_c}
- 2\, \frac{w+1}{w-1}\, \^0\tensor{\mathring{\sigma}}{^b_c} \,
t^{\textfrac{w-1}{w+1}} + O\big( t^{\textfrac{2(w-1)}{1+w}}\big) \right)
\:. \label{asybehgenwginv}
\end{align}
\end{subequations}
The expressions~\eqref{asybehgenw} replace~\eqref{asybeh} in
the context of the case of general $w
> 1$. It is not difficult to convince oneself
that~\eqref{asybehgenw} reduces to~\eqref{asybeh} in the
special case $w = 3$.

\section{Main result}\label{sec:mainresult}

Stated in an informal way, the main result we will prove is
that (`typical') solutions of the Einstein-Euler equations
`look like' the solutions~\eqref{asybehgenw} of the asymptotic
system in the asymptotic limit toward the singularity. The
rigorous statement is the following.

\begin{theorem}\label{thm2}
Let $\^0\mathcal{W} = \big(\^0g_{a b}, \tr \^0k,
\^0\tensor{\sigma}{^a_b}, \^0\mu,\^0u_a\big)$ be a solution of
the asymptotic system as given by~\eqref{asybehgenw}. Then
there exists a unique solution $\mathcal{W} = \big(g_{a b}, \tr
k, \tensor{\sigma}{^a_b}, \mu,u_a\big)$ of the Einstein-Euler
equations, whose asymptotic behavior, to leading order, is that
of $\^0\mathcal{W}$. Specifically, there exists a set of
quantities $\mathcal{U} = (\tensor{\gamma}{^a_b},\ \kappa,\
\tensor{\varsigma}{^a_b},\ \nu,\ v_a)$ that vanish in the limit
$t\rightarrow 0$, together with their spatial derivatives, and
an arbitrarily small $\varepsilon > 0$, such that
\begin{subequations}\label{result2}
\begin{align}
g_{ab} &= \^0g_{ab}  +t^{2 - 4/3(1+w) - 2 \varepsilon} \^0g_{ac}\,
\tensor{\gamma}{^c_b}\,, \\
\tr k &=   \tr\^0k  + t^{2 - 10/3(1+w) -\varepsilon}\:\kappa \,,\\
\tensor{\sigma}{^a_b} & =\^0\tensor{\sigma}{^a_b} + t^{1 -4/3(1+w)
-\varepsilon/2}\: \tensor{\varsigma}{^a_b}\,,\\
\mu &= \^0\mu \: + t^{1 - 10/3(1+w)-\varepsilon}\:\nu
\,, \\
\label{resultu2}
u_a &= \^0u_a  + t^{4-10/3(1+w)-2\varepsilon}\:v_a \:.
\end{align}
\end{subequations}
The inverse metric is given by $g^{ab} = \^0g^{ab} + t^{2 -
4/3(1+w) -2 \varepsilon}\,\bar{\gamma}^a{}_c\^0g^{cb}$, where
$\tensor{\bar{\gamma}}{^a_b}$ shares the properties of
$\mathcal{U}$.
\end{theorem}

\begin{remark}
The number of free functions (of the spatial variables) in the
set of solutions of the asymptotic system is the same as the
number of free functions that determine initial data for the
Einstein-Euler equations. Therefore, in the sense of function
counting, Theorem~\ref{thm2} shows that `typical' (`generic')
solutions of the Einstein-Euler system behave according
to~\eqref{result2}.
\end{remark}

\section{The reduced system}\label{sec:reducedsystem}

The proof of Theorem~\ref{thm2} is based on techniques in
connection with Fuchsian equations~\cite{Kichenassamy,
Andersson&Rendall, Kichenassamy/Rendall:1998}. Our aim is to
construct, from the Einstein-Euler evolution
equations~\eqref{evol} and~\eqref{euler} and the asymptotic
system~\mbox{\eqref{evoleqs}--\eqref{constraint equations}}, a system
of (generalized) Fuchsian type, i.e. a system of the form
\begin{equation}\label{fuchsian}
t\partial_t \mathcal{U} + \mathcal{A}\, \mathcal{U}  =
\mathcal{F}(t, x, \mathcal{U}, \partial_x\mathcal{U}) +
\mathcal{G}(t, x, \mathcal{U})\:t\,\partial_t \mathcal{U}
\end{equation}
where $\mathcal{U}$ denotes the collection of the variables of
Theorem~\ref{thm2} (and additional variables). In this context,
$\mathcal{A}$ is required to be a time-independent matrix that
depends analytically on the spatial coordinates, and
$\mathcal{F}$ and $\mathcal{G}$ are functions that are
continuous in $t > 0$ and analytic in the other variables.
Sufficient conditions that the system~\eqref{fuchsian} be a
(generalized) Fuchsian system are that (i) the matrix
$\mathcal{A}$ is the direct sum of a submatrix whose
eigenvalues have positive real parts and the zero matrix, and
that (ii) the functions $\mathcal{F}$ and $\mathcal{G}$ vanish
like a positive power of $t$ as $t\rightarrow 0$. If this is
the case, then there exists a unique solution $\mathcal{U}$
of~\eqref{fuchsian} such that $\mathcal{U}$ and its spatial
derivatives vanish with $t$; see~\cite{Kichenassamy,
Andersson&Rendall}. Our aim is to construct a
system~\eqref{fuchsian} and to show that the `Fuchsian
conditions' are met.

Let $\^0\mathcal{W} = \big(\^0g_{a b}, \^0g^{a b}, \tr \^0k,
\^0\tensor{\sigma}{^a_b}, \^0\mu,\^0u_a\big)$ be a solution of
the asymptotic system as given by~\eqref{asybehgenw}. Let us
(re)define $\mathcal{U}$ as $\mathcal{U} =
(\tensor{\gamma}{^a_b},\ \tensor{\bar{\gamma}}{^a_b},\ \kappa,\
\tensor{\varsigma}{^a_b},\ \nu,\ v_a,
\tensor{\lambda}{^a_b_c},\ \tensor{\bar{\lambda}}{^a_b_c})$. We
make the ansatz
\begin{subequations}\label{ansatz}
\begin{alignat}{2}
\label{ansatzg}
g_{ab} &= \^0g_{ab}  +t^{\alpha_\gamma} \^0g_{ac}\,
\tensor{\gamma}{^c_b}\,, & &\\
\label{ansatzginv}
g^{ab} &= \^0g^{ab} + t^{\bar{\alpha}_\gamma}\,
\bar{\gamma}^a{}_c\^0g^{cb}, & & \\
\tr k &= \tr\^0k + t^{\alpha_\kappa+1 -
\frac{4}{3(w+1)}}\:\kappa\,,\\
\tensor{\sigma}{^a_b} &= \^0\tensor{\sigma}{^a_b}  +
t^{-\alpha_\varsigma+1- \frac{4}{3(w+1)}}\:
\tensor{\varsigma}{^a_b}\,,\\
\mu &= \^0\mu\: + t^{\alpha_\nu - \frac{4}{3(w+1)}}\:\nu\,, \\
\label{ansatzu} u_a &= \^0u_a  + t^{2 w/(1+w) + \alpha_v}\:v_a\:,\\
\gamma^a{}_{b,c} &= t^{-\alpha_\lambda}\lambda^a_{bc}\:,\\
\bar{\gamma}^a{}_{b,c} &=
t^{-\bar{\alpha}_\lambda}\bar{\lambda}^a_{bc}\:,
\end{alignat}
\end{subequations}
where $(\alpha_\gamma, \bar{\alpha}_\gamma, \alpha_\kappa,
\alpha_\varsigma, \alpha_\nu, \alpha_v, \alpha_\lambda,
\bar{\alpha}_\lambda)$ are constants.

The procedure is (almost) straightforward: Imposing the
Einstein-Euler equations yields a system of equations for the
variables $\mathcal{U}$. The plan is then to find a range of
the free constants such that this system is Fuchsian, which in
turn implies the existence of a unique solution $\mathcal{U}$
with the boundary condition $\mathcal{U} \rightarrow 0$ (and
$\partial_x \mathcal{U} \rightarrow 0$) as $t\rightarrow 0$.

However, a slight modification of the Einstein-Euler equations
is needed for this purpose; cf.~\cite{Andersson&Rendall}. The
problem concerns the symmetry of the metric tensor $g_{a b}$.
Obviously, for an initial value problem, the issue of symmetry
does not arise: The Einstein evolution equations propagate the
symmetry of the metric (and the second fundamental form). In
the present context, however, the solution $\mathcal{U}$ of the
Fuchsian system we intend to obtain, is not determined by its
initial value but by requiring the boundary condition
$\mathcal{U} \rightarrow 0$ ($t\rightarrow 0$). It is thus not
clear, a priori, whether the tensor $g_{a b}$
of~\eqref{ansatzg} will be a symmetric tensor, since we do not
have control over the properties of $\tensor{\gamma}{^a_b}$.
This is turn suggests that special care is needed in connection
with the curvature terms in the Einstein equations:
In~\eqref{evol} and~\eqref{constr} we replace the curvature
terms $\tensor{R}{^a_b}$ and $R$ by the corresponding terms
that are defind from the symmetric part of the metric
${}^S\!g_{ab} = g_{(a b)}$; i.e., $\tensor{R}{^a_b}
\dashrightarrow {}^S\!\tensor{R}{^a_b} =
\tensor{R}{^a_b}[{}^S\!g_{cd}]$. Thereby we modify the
Einstein-Euler equations in such a way that the equations make
sense for general non-symmetric tensors but still reduce to the
original equations in case of symmetric data. In addition, we
must establish conventions concerning raising and lowering of
indices with a metric tensor that need not be symmetric. We
define $g^{ab}$ as the unique inverse of $g_{ab}$,
$g_{ac}g^{cb}= \tensor{\delta}{_a^b}$, and use the convention
that indices on tensors are raised and lowered with the second
index of $g^{a b}$ and $g_{a b}$, respectively. As a
consequence, index raising and lowering are inverse operations,
as is the case for symmetric metrics.

Having prepared the Einstein-Euler equations in this way we can
insert the ansatz~\eqref{ansatz} into~\eqref{evol}
and~\eqref{euler} without difficulty, and we obtain the
\emph{reduced system}, a system of equations for the variables
$\,\mathcal{U}$. (By construction, this system is well-defined
irrespective of the symmetry properties of the variables
$\,\mathcal{U}$. We will return to the question of symmetry of
the unique solution we subsequently construct in a separate
subsection.)

The reduced system reads
\begin{subequations}\label{reduced system}
\begin{align}
\nonumber
& t\partial_t\tensor{\gamma}{^a_b} + \alpha_\gamma\tensor{\gamma}{^a_b}
= 2\,t\,\big(\tensor{\gamma}{^a_c}\^0\tensor{\sigma}{^c_b} - \^0\tensor{\sigma}{^a_c}\tensor{\gamma}{^c_b} \big)
+ 2 \,t^{2(1+3 w)/3(1+w) -\alpha_\gamma -\alpha_\varsigma} \,\times \: \\[0.5ex]
\label{metriceq}
& \hspace{0.4\textwidth}\times\Big( \tensor{\varsigma}{^a_b}
+ t^{\alpha_\gamma} \,\tensor{\gamma}{^a_c} \tensor{\varsigma}{^c_b}
-\textfrac{1}{3}\, t^{\alpha_\kappa +\alpha_\varsigma}\,\kappa\, \big( \tensor{\delta}{^a_b}
+ t^{\alpha_\gamma} \,\tensor{\gamma}{^a_b} \big) \Big) \:,\\[1.5ex]
\nonumber
& t\partial_t\tensor{\bar{\gamma}}{^a_b} + \bar{\alpha}_\gamma
\tensor{\bar{\gamma}}{^a_b}
= 2\,t\,\big( \tensor{\bar{\gamma}}{^a_c}\^0\tensor{\sigma}{^c_b}
- \^0\tensor{\sigma}{^a_c}\tensor{\bar{\gamma}}{^c_b} \big)
- 2 \,t^{2(1+3 w)/3(1+w) -\bar{\alpha}_\gamma -\alpha_\varsigma}
\,\times \:\\[0.5ex]
\label{inversemetriceq}
& \hspace{0.4\textwidth}\times\Big(
\tensor{\varsigma}{^a_b}
+ t^{\alpha_\gamma} \,\tensor{\varsigma}{^a_c} \tensor{\bar{\gamma}}{^c_b}
-\textfrac{1}{3}\, t^{\alpha_\kappa +\alpha_\varsigma}\,\kappa\,
\big( \tensor{\delta}{^a_b}
+ t^{\bar{\alpha}_\gamma} \,\tensor{\bar{\gamma}}{^a_b} \big) \Big)
\:,\\[1.5ex]
\nonumber
& t\partial_t\kappa + (1 + \alpha_\kappa)\kappa - \textfrac{1}{2}
(1+3 w) t^{\alpha_\nu-\alpha_\kappa} \,\nu  =
2 t^{1-\alpha_\kappa-\alpha_\varsigma}\,\tensor{\varsigma}{^a_b}
\^0\tensor{\sigma}{^b_a}
+ \textfrac{4}{3(1+w)} \, \big[ 1 + \textfrac{1+w}{2} \,t \,\tr\^0k
\big] \,\kappa \\[0.5ex]
\nonumber
& \hspace{0.4\textwidth} + t^{2(1+3 w)/3(1+w)-\alpha_\kappa-
2\alpha_\varsigma} \,
\big( \tensor{\varsigma}{^a_b} \tensor{\varsigma}{^b_a}+
\textfrac{1}{3}\, t^{2 \alpha_\kappa+2 \alpha_\varsigma}\,
\kappa^2\big) \\[0.5ex]
\label{mean curvatureeq}
& \hspace{0.4\textwidth} + (1+w) t^{4/3(1+w) - \alpha_\kappa} \,
\^0\mu\, u^2 + (1+w) t^{\alpha_\nu-\alpha_\kappa} u^2\, \nu\,,
\\[1.5ex]
\nonumber
& t\partial_t \tensor{\varsigma}{^a_b}  + \big(1+\textfrac{2}
{3(1+w)}- \alpha_\varsigma) \tensor{\varsigma}{^a_b}  =
t^{1 + \alpha_\kappa + \alpha_\varsigma}\^0\sigma^a{}_b \,\kappa
+ t^{2(1+3 w)/3(1+w) + \alpha_\kappa}\varsigma^a{}_b\,\kappa \\
\nonumber
& \hspace{0.33\textwidth} +\textfrac{2}{1+w}  \, \big[ 1 +
\textfrac{1+w}{2} \,t \,\tr\^0k \big]\,\tensor{\varsigma}{^a_b}
- t^{4/3(1+w) + \alpha_\varsigma}\big({}^S\!\tensor{R}{^a_b}-
\textfrac{1}{3} {}^S\!R \,\tensor{\delta}{^a_b}\big) \\[0.5ex]
\nonumber
& \hspace{0.33\textwidth} + (1+w) t^{4/3(1+w) + \alpha_\varsigma}
\^0\mu \big( u^a u_b - \textfrac{1}{3} \,u^2 \tensor{\delta}{^a_b}
\big) \\[0.5ex]
\label{sheareq}
& \hspace{0.33\textwidth}
+ (1+w) t^{\alpha_\nu + \alpha_\varsigma} \big( u^a u_b -
\textfrac{1}{3} \,u^2 \tensor{\delta}{^a_b} \big) \nu \,,\\[1.5ex]
\nonumber
& t\partial_t\nu + (2- \textfrac{4}{3(1+w)} + \alpha_\nu) \nu
- \textfrac{4}{3(1+w)} \,t^{\alpha_\kappa-\alpha_\nu} \,\kappa  =
2 \big[1 + \textfrac{1+w}{2}\,  t\, \tr\^0k \big] \nu + (1+w)
t^{2(1+3 w)/3(1+w) + \alpha_\kappa} \kappa \nu \\[0.5ex]
\nonumber
& \hspace{0.4\textwidth}
- \textfrac{4}{3(1+w)} \,t^{\alpha_\kappa-\alpha_\nu} \big[1 -
\textfrac{3}{4} (1+w)^2 \^0\mu \:t^2\big] \kappa \\[0.5ex]
\label{energyeq}
& \hspace{0.4\textwidth}
+ t^{1+4/3(1+w)-\alpha_\nu} \big[\text{r.h.s.\
of~\eqref{eulermu}}\big] \,,\\[1.5ex]
\nonumber
& t\partial_tv_a + \alpha_vv_a =
-\textfrac{2 w}{1+w} \big[1 + \textfrac{1+w}{2}\,  t\, \tr\^0k
\big] \,v_a
- w \,t^{2/3(1+w) + \alpha_\kappa-\alpha_v} \kappa \^0u_a  -w
t^{2(1+3 w)/3(1+w) + \alpha_\kappa} \kappa v_a  \\[0.5ex]
\nonumber
& \hspace{0.25\textwidth}
-\textfrac{w}{1+w} (\^0\mu)^{-1} t^{-1 + 2/3(1+w) + \alpha_\nu
- \alpha_v} \nabla_a \nu
+\textfrac{w}{1+w} (\^0\mu)^{-1} t^{-1 + 2/3(1+w) + \alpha_\nu
- \alpha_v}\,\times
\\[0.5ex]
\nonumber
& \hspace{0.3\textwidth}
\times
\big( \^0\mu + t^{-4/3(1+w) + \alpha_\nu} \nu \big)^{-1} \big(
\nabla_a \^0\mu + t^{-4/3(1+w) + \alpha_\nu} \nabla_a \nu\big)\,\nu
\\[0.5ex]
\label{velocityeq}
& \hspace{0.25\textwidth}
+ t^{1-2 w/(1+w)-\alpha_v} \big[\text{r.h.s.\ of~\eqref{euleru}}\big]
\,,\\[1.5ex]
\label{metricgradeq1}
& t\partial_t \lambda^a_{bc}  = t^{\alpha_\lambda}t\partial_t
\gamma^a{}_{b,c} + \alpha_\lambda t^{\alpha_\lambda} \gamma^a{}_{b,c}\,,
\\[0.8ex]
\label{metricgradeq2}
& t\partial_t \bar{\lambda}^a_{bc}  = t^{\bar{\alpha}_\lambda}
t\partial_t\bar{\gamma}^a{}_{b,c} + \bar{\alpha}_\lambda
t^{\bar{\alpha}_\lambda} \bar{\gamma}^a{}_{b,c}.
\end{align}
\end{subequations}

For brevity we do not write out explicitly the terms
on the right hand sides in equations~\eqref{energyeq}
and~\eqref{velocityeq}. The expressions in question are
straightforwardly obtained by inserting the
ansatz~\eqref{ansatz} into the r.h.s.~of~\eqref{euler}.

\section{Fuchsian analysis} \label{sec:Fuchsiananalysisgenw}

\subsection{The reduced system as a Fuchsian system}

In this section we perform the proof of Theorem~\ref{thm2}.
The first part of the proof of is to show that there exist appropriate
choices of the constants $(\alpha_\gamma, \bar{\alpha}_\gamma,
\alpha_\kappa, \alpha_\varsigma, \alpha_\nu, \alpha_v,
\alpha_\lambda, \bar{\alpha}_\lambda)$ such that the reduced
system~\eqref{reduced system} becomes a Fuchsian system.

The first step is to consider the l.h.s.\ of~\eqref{reduced
system}. As a prerequisite, the l.h.s.\ must be of the form
$t\partial_t \mathcal{U} + \mathcal{A}\, \mathcal{U}$, where
$\mathcal{U}$ denotes the collection of variables, i.e.,
$\mathcal{U} = (\tensor{\gamma}{^a_b},\
\tensor{\bar{\gamma}}{^a_b},\ \kappa,\
\tensor{\varsigma}{^a_b},\ \nu,\ v_a,
\tensor{\lambda}{^a_b_c},\ \tensor{\bar{\lambda}}{^a_b_c})$.
This requires setting
\[
\alpha_\kappa = \alpha_\nu =: \beta\:.
\]
To establish the conditions on the coefficient matrix
$\mathcal{A}$ that are required for the system to be Fuchsian,
cf.~\cite[Section 4]{Andersson&Rendall}, we show that
$\mathcal{A}$ is the direct sum of a positive definite matrix
and the zero matrix, provided that the constants
$\big(\alpha_\gamma, \bar{\alpha}_\gamma, \alpha_\kappa [=
\beta], \alpha_\varsigma, \alpha_\nu [= \beta], \alpha_v,
\alpha_\lambda, \bar{\alpha}_\lambda\big)$ are chosen
appropriately.

Obviously, the only non-diagonal part of $\mathcal{A}$,
associated with the variables $(\kappa, \nu)$, corresponds to
the submatrix
\begin{displaymath}
\begin{pmatrix}
1+ \beta & -\textfrac{1}{2} (1 + 3 w) \\[0.5ex]
-\textfrac{4}{3} (1+ w)^{-1} & 2 -\textfrac{4}{3} (1+ w)^{-1} +
\beta
\end{pmatrix}\,.
\end{displaymath}
The eigenvalues of this matrix are $\beta$ and $3 -
\textfrac{4}{3} (1+w)^{-1} + \beta$, which are positive if and
only if $\beta > 0$. Therefore, the conditions ensuring
positive definiteness of the submatrix associated with
$\big(\tensor{\gamma}{^a_b}, \tensor{\bar{\gamma}}{^a_b},
\kappa, \tensor{\varsigma}{^a_b}, \nu, v_a\big)$ are
\begin{equation}\label{cond1genw}
\mathrm{a})\quad \alpha_\gamma > 0\qquad
\mathrm{b})\quad \bar{\alpha}_\gamma > 0\qquad
\mathrm{c})\quad \alpha_\kappa = \alpha_\nu = \beta > 0 \qquad
\mathrm{d})\quad \alpha_\varsigma < 1 + \frac{2}{3(1+w)}\qquad
\mathrm{e})\quad \alpha_v > 0 \:.
\end{equation}

In the second step we turn to the r.h.s.\ of~\eqref{reduced
system}, which is of the form $\mathcal{F} +\mathcal{G}\, t
\partial_t \mathcal{U}$, where $\mathcal{F}$ depends on $t$,
the spatial variables, and $\mathcal{U}$ and its first spatial
derivatives, and $\mathcal{G} = \mathcal{G}(t,x,\mathcal{U})$.
We need to show that $\mathcal{F}$ and $\mathcal{G}$ $\in
O(t^\delta)$ for some positive number $\delta$, if the
constants $\big(\alpha_\gamma, \bar{\alpha}_\gamma,
\alpha_\kappa [= \beta], \alpha_\varsigma, \alpha_\nu [=
\beta], \alpha_v, \alpha_\lambda, \bar{\alpha}_\lambda\big)$
are chosen appropriately.
We use the notation $f(t,x) \in
O\big(g(t)\big)$ to compare the asymptotic
behavior of a function $f(t,x)$ and a function $g(t)$ near $t=0$, if
$f(t,x) = O\big( g(t) \big)$ as $t\rightarrow 0$ uniformly on compacts
subsets of the spatial variables $x$.

Equations~\eqref{metriceq} and~\eqref{inversemetriceq} imply
the conditions
\begin{equation}\label{cond2genw}
\mathrm{a})\quad \alpha_\gamma + \alpha_\varsigma <
\frac{2(1+3 w)}{3(1+w)}\qquad
\mathrm{b})\quad \bar{\alpha}_\gamma + \alpha_\varsigma <
\frac{2(1+3 w)}{3(1+w)}\:.
\end{equation}
To treat~\eqref{mean curvatureeq} we use that $u^2$ differs
from ${}^0u^2$ by a function of time that goes to zero as
$t\rightarrow 0$ faster than ${}^0u^2$ itself,
cf.~\eqref{ansatzginv},~\eqref{ansatzu},
and~(\ref{cond1genw}b,e). The leading order is thus obtained by
replacing $u^2$ by ${}^0u^2$ in~\eqref{mean curvatureeq}. We
get the conditions
\begin{equation}\label{cond3genw}
\mathrm{a})\quad \alpha_\zeta + \beta < 1 - \frac{2}{1+w}
\qquad \mathrm{b})\quad
2 \alpha_\zeta + \beta<  2 - \frac{4}{3(1+w)}
\qquad \mathrm{c})\quad
\beta < 2 -  \frac{4}{1+w}\:.
\end{equation}
Analogously,~\eqref{sheareq} leads to
\begin{equation}\label{cond4genw}
\mathrm{a})\quad \alpha_\zeta + \beta > -\big( 1 - \frac{2}{1+w}\big) =
\frac{1-w}{1+w}\qquad \mathrm{b})\quad
\alpha_\zeta > -\big(  2 -  \frac{4}{1+w}\big) = 2\, \frac{1-w}{1+w}\:;
\end{equation}
the estimate of the curvature term in~\eqref{sheareq} is not
included in these conditions; we refer to
Corollary~\ref{curvcor}.

The analysis of~\eqref{energyeq} requires a thorough
investigation of the terms on the r.h.s.\ of~\eqref{eulermu}.
In this process, we may replace, in each term, the variables
$\big(g_{a b}, g^{a b}, \tr k, \tensor{\sigma}{^a_b},
\mu,u_a\big)$ by the solution of the asymptotic system
$\big(\^0g_{a b}, \^0g^{a b}, \tr \^0k,
\^0\tensor{\sigma}{^a_b}, \^0\mu,\^0u_a\big)$. To see that this
replacement is possible, consider, e.g., the term $\mu u^a
\partial_t u_a$ of~\eqref{eulermu}. Inserting~\eqref{ansatz}
the term $\mu u^a \partial_t u_a$ becomes
\[
\big( \^0\mu + t^{\beta-4/3(1+w)} \nu \big)
\big( \^0g^{a b} + t^{\bar{\alpha}_\gamma} \tensor{\bar{\gamma}}{^a_c}  \^0g^{b c} \big)
\big( \^0u_b + t^{2 w/(1+w) + \alpha_v} v_b \big)
\big( \partial_t\^0u_a + t^{2 w/(1+w) + \alpha_v-1} [\mathrm{const}\; v_a + t \partial_t v_a ] \big)\:.
\]
Because of~(\ref{cond1genw}e), the power $t^{2 w/(1+w) +
\alpha_v}$ goes to zero as $t\rightarrow 0$ faster than
${}^0u_a$; the same is true for $t^{2 w/(1+w) + \alpha_v-1}$ in
comparison with $\partial_t\^0u_a$ and $t^{\beta-4/3(1+w)}$ in
comparison with ${}^0\mu$. We conclude that the term
$t^{1+4/3(1+w) -\beta} \,\mu u^a \partial_t u_a $
of~\eqref{energyeq} is of the required form $\mathcal{F}
+\mathcal{G}\, t \partial_t \mathcal{U}$, where $\mathcal{F}$
and $\mathcal{G}$ converge to zero as $O(t^\delta)$ for some
positive power $\delta$, if $t^{1+4/3(1+w) -\beta} \^0\mu
\^0u^a\partial_t\!\^0u_a \in O(t^\delta)$.

We find that the term on the r.h.s.\ of~\eqref{energyeq} with
the slowest convergence is $t^{1+4/3(1+w) -\beta} \mu
\sqrt{1+u^2} \nabla_a u^a$. This term thus yields the most
restrictive condition on $\beta$, which reads
\begin{equation}\label{cond5genw}
\beta < 1 - \frac{2}{1+w} = \frac{w-1}{w+1}\:.
\end{equation}

Finally, in order for the r.h.s.\ of~\eqref{velocityeq} to be
of the required form, we must impose the additional conditions
\begin{equation}\label{cond6genw}
\mathrm{a})\quad \alpha_v < 1 + \frac{2}{3(1+w)} + \beta =
\frac{3w+5}{3(1+w)} + \beta \qquad \mathrm{b})\quad
\alpha_v < 3 - \frac{10}{3(1+w)} = \frac{9w-1}{3(1+w)}\:.
\end{equation}
Equations~\eqref{metricgradeq1} and~\eqref{metricgradeq2} yield
$\alpha_\lambda >0$ and $\bar{\alpha}_\lambda > 0$.

In the remainder of this subsection we consider the most
complicated expressions of the r.h.s.\ of~\eqref{reduced
system}, which are the curvature terms in~\eqref{sheareq}.
The estimates of the curvature terms are presented in a
succession of lemmas. Throughout we use a coordinate frame to
perform the computations.

\begin{lemma}
\begin{equation}
\^0g_{ab}, \^0g_{ab,c}, \^0g_{ab,cd} \in O(t^{\frac{4}{3(1+w)}}),
\quad \^0g^{ab}, \^0g^{ab}{}_{,c} \in
O(t^{-\frac{4}{3(1+w)}}).
\end{equation}
\end{lemma}

\begin{proof}
The statements follow directly from the series expansions
equations~\eqref{asybehgenwg},~\eqref{asybehgenwginv} and the
smoothness of the initial data.
\end{proof}

\begin{lemma}\label{Curvature estimate 1}
\begin{equation}
{}^{0}\!R_{ab} \in O(1).
\end{equation}
\end{lemma}

\begin{proof}
In a coordinate frame we have
\begin{align*}
2{}^{0}\Gamma^a{}_{bc} & = \underbrace{\^0g^{ad}}_{O(t^{-\frac{4}{3(1+w)}})}
\underbrace{(\^0g_{db,c} + \^0g_{dc,b} - \^0g_{bc,d})}
_{=:\^0\Gamma_{abc} \in O(t^{\frac{4}{3(1+w)}})} \in O(1)\:,\\
2{}^{0}\Gamma^a{}_{bc,d} =& \underbrace{\^0g^{ae}{}_{,d}}
_{O(t^{-\frac{4}{3(1+w)}})}\underbrace{\^0\Gamma_{ebc}}
_{O(t^{\frac{4}{3(1+w)}})} + \underbrace{\^0g^{ae}}
_{O(t^{-\frac{4}{3(1+w)}})}\underbrace{\^0\Gamma_{ebc,d}}
_{O(t^{\frac{4}{3(1+w)}})} \in O(1),
\end{align*}
which gives $\^0R_{ab} = {}^{0}\Gamma^c{}_{ab,c} -
{}^{0}\Gamma^c{}_{ac,b} +
{}^{0}\Gamma^c{}_{dc}{}^{0}\Gamma^d{}_{ab} -
{}^{0}\Gamma^c{}_{db}{}^{0}\Gamma^d{}_{ac} \in O(1)$.
\end{proof}

\begin{lemma}\label{Curvature estimate 2}
If $\alpha_\gamma > \alpha_\lambda > 0$, and
$\bar{\alpha}_\gamma
> \bar{\alpha}_\lambda > 0$ then
\begin{equation}
{}^S\!R_{ab} \in O(1).
\end{equation}
\end{lemma}

\begin{proof}
Assume $\alpha_\gamma > \alpha_\lambda > 0$, and
$\bar{\alpha}_\gamma
> \bar{\alpha}_\lambda > 0$. We have
\begin{align*}
{}^Sg_{ab} &= \^0g_{ab} + \frac{1}{2}t^{\alpha_\gamma}(\^0g_{ac}
\gamma^c{}_b+ \^0g_{bc}\gamma^{c}{}_a) \in O(\^0g_{ab})
 \subseteq O(t^{\frac{4}{3(1+w)}}),\\[1ex]
{}^Sg^{ab} & = \^0g^{ab} + t^{\bar{\alpha}_\gamma}
\frac{1}{2}(\bar{\gamma}^a{}_c\^0g^{cb} + \bar{\gamma}^b{}_c\^0g^{ca})
\in O(\^0g^{ab}) \subseteq O(t^{-\frac{4}{3(1+w)}}),
\end{align*}
\begin{align*}
{}^Sg_{ab,c} & = \^0g_{ab,c} + \frac{1}{2}t^{\alpha_\gamma}(\^0g_{ad,c}
\gamma^d{}_b+ \^0g_{bd,c}\gamma^{d}{}_a) + \frac{1}{2}t^{\alpha_\gamma
- \alpha_\lambda}(\^0g_{ad}\lambda^d{}_{bc}+ \^0g_{bd}\lambda^{d}
{}_{ac}) \in O(t^{\frac{4}{3(1+w)}}), \\[1ex]
{}^Sg^{ab}{}_{,c} & = \^0g^{ab}{}_{,c} + \frac{1}{2}t^{\bar{\alpha}_\gamma}
(\bar{\gamma}^a{}_d\^0g^{db}{}_{,c} + \bar{\gamma}^b{}_d\^0g^{da}
{}_{,c})+ \frac{1}{2}t^{\bar{\alpha}_\gamma
- \bar{\alpha}_\lambda}(\bar{\lambda}^a{}_{dc}\^0g^{db}
+ \bar{\lambda}^b{}_{dc}\^0g^{da}) \in O(t^{-\frac{4}{3(1+w)}}),
\end{align*}
\begin{align*}
{}^Sg_{ab,cd} = \^0g_{ab,cd} +
\frac{1}{2}t^{\alpha_\gamma}(\^0g_{ae,cd} \gamma^e{}_b+
\^0g_{be,cd}\gamma^{e}{}_a) & + t^{\alpha_\gamma -
\alpha_\lambda}(\^0g_{ae}\lambda^e{}_{bc}+ \^0g_{be}\lambda^{e}
{}_{ac}) \\
& + \frac{1}{2}t^{\alpha_\gamma -
\alpha_\lambda}(\^0g_{ae}\lambda^e{}_{bc,d}+
\^0g_{be}\lambda^{e} {}_{ac,d}) \in O(t^{\frac{4}{3(1+w)}})\:.
\end{align*}
The result then follows as in the proof of Lemma \ref{Curvature
estimate 1}.
\end{proof}

\begin{corollary}\label{curvcor}
The curvature term $t^{4/3(1+w)
+\alpha_\varsigma}\big({}^S\!\tensor{R}{^a_b}- \textfrac{1}{3}
{}^S\!R \,\tensor{\delta}{^a_b}\big)$ on the r.h.s.\ of
equation~\eqref{sheareq} vanishes with $t$ if
$\alpha_\varsigma$ is any positive constant and the conditions
of Lemma \ref{Curvature estimate 2} are met.
\end{corollary}

It remains to collect
conditions~\eqref{cond1genw}--\eqref{cond6genw} and the result
of Corollary~\ref{curvcor}: The reduced system~\eqref{reduced
system} is a Fuchsian system of the form~\eqref{fuchsian} that
satisfies the requirements on the matrix and the functions if
the constants $\big(\alpha_\gamma, \bar{\alpha}_\gamma,
\alpha_\kappa [= \beta], \alpha_\varsigma, \alpha_\nu [=
\beta], \alpha_v, \alpha_\lambda, \bar{\alpha}_\lambda\big)$
are chosen according to the conditions
\begin{equation}\label{condisag}
\begin{split}
& 0 < \alpha_\lambda < \alpha_\gamma \:,\qquad
0 < \bar{\alpha}_\lambda < \bar{\alpha}_\gamma  \:,\qquad
\alpha_\gamma + \alpha_\varsigma < 2 - \frac{4}{3(1+w)} \:,\qquad
\bar{\alpha}_\gamma + \alpha_\varsigma < 2 - \frac{4}{3(1+w)} \:,\\
& \beta < 1 - \frac{2}{1+w}\:,\qquad
\alpha_\varsigma + \beta <  1 - \frac{2}{1+w} \:,\qquad
\alpha_v < 1 + \frac{2}{3(1+w)} + \beta \:.
\end{split}
\end{equation}
Under these conditions, there exists, for each choice of
solution $\^0\mathcal{W} = \big(\^0g_{a b}, \tr \^0k,
\^0\tensor{\sigma}{^a_b}, \^0\mu,\^0u_a\big)$ of the asymptotic
system, a unique solution $\mathcal{U}$ of the reduced system
that vanishes as $t\rightarrow 0$.

To obtain the statement~\eqref{result2} of Theorem~\ref{thm2}
we set
\begin{equation}
\alpha_\gamma = \bar{\alpha}_\gamma =  2 - \frac{4}{3(1+w)} -2 \varepsilon\:,\quad
\alpha_\kappa = \alpha_\nu  = \beta =  1 - \frac{2}{1+w} - \varepsilon \:,\quad
\alpha_\varsigma = \frac{\varepsilon}{2} \:,\quad
\alpha_v = 2 -  \frac{4}{3(1+w)} -  2 \varepsilon
\end{equation}
which is compatible with~\eqref{condisag} for arbitrarily small
$\varepsilon > 0$, and insert these constants
into~\eqref{ansatz}.

\subsection{Symmetry of the solution of the Fuchsian system}

In the second part of the proof of Theorem~\ref{thm2} we need
to come back to the problem of symmetry: Using the reduced
system and its unique solutions that vanish as $t\rightarrow
0$, we have constructed solutions \mbox{$\mathcal{W} =
\big(g_{a b}, \tr k, \tensor{\sigma}{^a_b}, \mu,u_a\big)$} of
the Einstein-Euler evolution equations~\eqref{evol}
and~\eqref{euler} with modified curvature terms,
${}^S\!\tensor{R}{^a_b}$ and ${}^S\!R$, cf.~the remarks
following~\eqref{ansatz}. A priori, it is not obvious that the
constructed tensors $g_{a b}$ and $\sigma_{a b}$ are symmetric
tensors; however, if they are, then the solutions $\mathcal{W}$
we have obtained are solutions of the proper Einstein-Euler
evolution equations. (The Einstein-Euler constraint equations
are analyzed in the subsequent subsection.)

Let $\^0\mathcal{W} = \big(\^0g_{a b}, \tr \^0k,
\^0\tensor{\sigma}{^a_b}, \^0\mu,\^0u_a\big)$ be a solution of
the asymptotic system as given by~\eqref{asybeh}. Consider the
unique solution $\mathcal{U}$ of the reduced
system~\eqref{reduced system} that vanishes with $t$ and
define, through~\eqref{ansatz}, the quantities $\mathcal{W} =
\big(g_{a b}, \tr k, \tensor{\sigma}{^a_b}, \mu,u_a\big)$. By
construction, $\mathcal{W}$ satisfies the Einstein-Euler evolution
equations with modified curvature terms. These equations imply
equations for the antisymmetric parts of $g_{a b}$, $g^{a b}$,
and $\sigma_{a b}$. We have
\begin{subequations}\label{symmetry eqs}
\begin{align}
\partial_t g_{[ab]}&= 2\sigma_{[ab]} - \textfrac{2}{3}(\tr k) g_{[ab]}
\,, \\[0.5ex]
\label{symm12}
\partial_t g^{[ab]}&= -2\sigma^{[a}{}_{c}g^{\hat{c} b]} + \textfrac{2}{3}(\tr k) g^{[ab]}
\,, \\[0.5ex]
\label{symm2}
\partial_t \sigma_{[ab]}&= \textfrac{1}{3}(\tr k)\sigma_{[ab]} +
2 \sigma_{[a\hat{c}}\sigma^c{}_{b]} + \textfrac{1}{3}(w+1)\mu u^2
g_{[ab]} + \textfrac{1}{3}{}^S\!R \:g_{[ab]} .
\end{align}
The terms $\sigma^{[a}{}_{c}g^{\hat{c} b]}$ and
$\sigma_{[a\hat{c}}\sigma^c{}_{b]}$ in~\eqref{symm12}
and~\eqref{symm2} can be expanded by splitting both
$\sigma_{ab}$ and $g^{ab}$ into its symmetric and antisymmetric
part. Each term in this expansion then contains at least one of
the anti-symmetric tensors $\sigma_{[a b]}$ or $g^{[a b]}$. We
have
\begin{align}\label{symmtermsexp}
\sigma^{[a}{}_{c}g^{\hat{c} b]} & =
g^{(a c)} \sigma_{(c d)} g^{[d b]} +
g^{[a c]} \sigma_{(c d)} g^{(d b)} +
g^{[a c]} \sigma_{[c d]} g^{[d b]} +
g^{(a c)}  g^{(b d)} \sigma_{[c d]}\:,\\[0.5ex]
\label{symmtermsexp2}
\sigma_{[a\hat{c}}\sigma^c{}_{b]} & =
\sigma_{(a c)} g^{(c d)} \sigma_{[d b]} +
\sigma_{[a c]} g^{(c d)} \sigma_{(d b)} +
\sigma_{[a c]} g^{[c d]} \sigma_{[d b]} +
\sigma_{(a c)} \sigma_{(b d)} g^{[c d]} \:.
\end{align}
\end{subequations}

Let $g$, $\bar{g}$, and $\sigma$ be constants and define the
antisymmetric tensors $\Omega^{\mathrm{g}}_{\,ab} :=
t^gg_{[ab]}$, $\Omega_{\bar{\mathrm{g}}}^{\,ab} := t^{\bar{g}}
g^{[ab]}$, and $\Omega^\sigma_{\,ab} := t^\sigma
\sigma_{[ab]}$. From~\eqref{symmetry eqs} we obtain a system of
equations for these tensors,
\begin{subequations}\label{Fuchsian symmetry eqs}
\begin{align}
\label{omegag}
t\partial_t \Omega^{\mathrm{g}}_{\, a b} + \big({-g}-\textfrac{4}{3}\textfrac{1}{1+w}\big)
\Omega^{\mathrm{g}}_{\,a b} &= 2t^{1+g-\sigma}\Omega^\sigma_{\,a b} - \textfrac{4}{3(1+w)}
\big[ 1 + \textfrac{1+w}{2} \,t \:\tr k\big] \Omega^{\mathrm{g}}_{\,a b}\:, \\[1ex]
\label{omega2}
t\partial_t \Omega_{\bar{\mathrm{g}}}^{\,a b} + \big({-\bar{g}}+\textfrac{4}{3}\textfrac{1}{1+w}\big)
\Omega_{\bar{\mathrm{g}}}^{\, a b} &= -2 t^{1+\bar{g}}
\sigma^{[a}{}_{c}g^{\hat{c} b]} + \textfrac{4}{3(1+w)}
\big[ 1 + \textfrac{1+w}{2} \,t \:\tr k\big] \Omega_{\bar{\mathrm{g}}}^{\, ab}\:, \\[1ex]
\nonumber
t\partial_t \Omega^\sigma_{ab} + \big({-\sigma} +\textfrac{2}{3}
\textfrac{1}{1+w} \big) \Omega^\sigma_{ab} &= 2t^{1+\sigma}
\sigma_{[a\hat{c}}\sigma^c{}_{b]} + \textfrac{2}{3 (1+w)} \big[ 1+ \textfrac{1+w}{2} \,t\: \tr k\big] \Omega^\sigma_{ab}  \\[0.5ex]
\label{omega3}
& \qquad + \textfrac{1}{3}t^{1+\sigma -g}
\:{}^S\!R\:\Omega^{\mathrm{g}}_{ab} - \textfrac{1}{3}(1+w)\mu u^2t^{1+\sigma -g}
\Omega^{\mathrm{g}}_{ab}\:,
\end{align}
where we use~\eqref{symmtermsexp} and~\eqref{symmtermsexp2} to express
$t^{1+\bar{g}} \,\sigma^{[a}{}_{c}g^{\hat{c} b]}$ and
$t^{1+\sigma} \sigma_{[a\hat{c}}\sigma^c{}_{b]}$
in terms of
$\Omega^{\mathrm{g}}_{\,ab}$, $\Omega_{\bar{\mathrm{g}}}^{\,ab}$, and $\Omega^\sigma_{\,ab}$,
according to
\begin{align}
\label{bterms}
t^{1+\bar{g}} \,\sigma^{[a}{}_{c}g^{\hat{c} b]} & =
 \underbrace{t g^{(a c)} \sigma_{(c d)}}_{\in O(t^{\frac{w-1}{w+1}})}  \Omega_{\bar{\mathrm{g}}}^{\, d b} +
 \underbrace{t \sigma_{(c d)} g^{(d b)}}_{\in O(t^{\frac{w-1}{w+1}})}  \Omega_{\bar{\mathrm{g}}}^{\, a c} +
 t^{1- \sigma - \bar{g}} \Omega_{\bar{\mathrm{g}}}^{\, a c} \Omega^\sigma_{\,c d}\Omega_{\bar{\mathrm{g}}}^{\,d b} +
 \underbrace{t^{1 +\bar{g} - \sigma}  g^{(a c)} g^{(b d)}}_{\in  O(t^{1 + \bar{g} - \sigma -
\frac{8}{3(1+w)}})} \Omega^\sigma_{\,c d} \\[1ex]
\label{aterms}
t^{1+\sigma} \sigma_{[a\hat{c}}\sigma^c{}_{b]} & =
 \underbrace{t\sigma_{(a c)} g^{(c d)}}_{\in O(t^{\frac{w-1}{w+1}})} \Omega^\sigma_{\,d b} +
 \underbrace{t g^{(c d)} \sigma_{(d b)}}_{\in O(t^{\frac{w-1}{w+1}})} \Omega^\sigma_{\,a c}  +
 t^{1 -\sigma -\bar{g}} \Omega^\sigma_{\,a c} \Omega_{\bar{\mathrm{g}}}^{\,c d} \Omega^\sigma_{\,d b}  +
 \underbrace{t^{1 + \sigma - \bar{g}} \sigma_{(a c)}\sigma_{(b d)}}_{\in O(t^{1  + \sigma - \bar{g}- \frac{4}{3(1+w)}})}
 \Omega_{\bar{\mathrm{g}}}^{\,c d} \:.
\end{align}
\end{subequations}

By an appropriate choice of the constants $g$, $\bar{g}$, and
$\sigma$ we are able to achieve that the system~\eqref{Fuchsian
symmetry eqs} is of the Fuchsian form~\eqref{fuchsian}, i.e.,
(i) the factors in front of $\Omega^{\mathrm{g}}_{\,ab}$,
$\Omega_{\bar{\mathrm{g}}}^{\,a b}$, and $\Omega^\sigma_{\,ab}$
on the l.h.\ sides of~\eqref{Fuchsian symmetry eqs} are
positive, and (ii) the factors of $\Omega^{\mathrm{g}}_{\,ab}$,
$\Omega_{\bar{\mathrm{g}}}^{\,a b}$, and $\Omega^\sigma_{\,ab}$ on 
the r.h.\ side vanish like a
power of $t$ with $t\rightarrow 0$. (As we have established,
the leading asymptotic behavior as $t\rightarrow 0$ of the
terms on the r.h.s.\ of~\eqref{Fuchsian symmetry eqs} is given
by replacing $g_{(a b)}$ by $\^0g_{a b}$ and
using~\eqref{asybehgenw}, and analogously for the remaining
expressions.) From the l.h.s.\ of~\eqref{Fuchsian symmetry eqs}
we have
\begin{subequations}\label{condisw}
\begin{equation}
g < -\frac{4}{3(1+w)} \:, \qquad \bar{g} < \frac{4}{3(1+w)} \:,\qquad
\sigma < \frac{2}{3(1+w)} \:.
\end{equation}
From the r.h.s.\ of equations~\eqref{omegag}--\eqref{aterms} we
obtain the conditions
\begin{align}
& 1 + g -\sigma > 0 \:,\qquad
1 - \sigma - \bar{g} > 0 \:, \qquad
1 + \bar{g} - \sigma - \frac{8}{3(1+w)} > 0 \:, \\
& 1  + \sigma - \bar{g}- \frac{4}{3(1+w)} > 0\:, \qquad
1  + \sigma  - g - \frac{4}{3(1+w)}> 0 \:.
\end{align}

The third set of conditions we impose on the constants $g$,
$\bar{g}$, and $\sigma$ is that $\Omega^{\mathrm{g}}_{\,ab}$,
$\Omega_{\bar{\mathrm{g}}}^{\,a b}$, and $\Omega^\sigma_{\,ab}$
vanish in the limit $t\rightarrow 0$. E.g.,
from~\eqref{asybehgenw} and~\eqref{result2} we have
\[
\Omega^{\mathrm{g}}_{\,a b} = t^{g + \alpha_\gamma} \^0g_{
[a \hat{c}} \gamma^c{}_{b]}\, \in \, O(t^{g + 2 -2 \varepsilon}) \:,
\]
while it turns out that $\Omega_{\bar{\mathrm{g}}}^{\,a b} \in
O(t^{\bar{g} + \frac{2}{3}\frac{3w-1}{1+w} - 2 \varepsilon})$
and $\Omega^\sigma_{\,a b} \in O(t^{\sigma + 1
-\varepsilon/2})$. This gives
\begin{equation}
g + 2 - 2 \varepsilon > 0 \:,\qquad
\bar{g} + 2 - \frac{8}{3(1+w)} - 2 \varepsilon > 0 \:, \qquad
\sigma + 1 - \frac{\varepsilon}{2} > 0 \:.
\end{equation}
\end{subequations}
The positive constant $\varepsilon$ can always be chosen small
enough so that it is possible to satisfy the collection of
conditions~\eqref{condisw} for an arbitrary $w > 1$. An
explicit choice is
\[
g = -\frac{5 + 3w}{3(1+w)}, \qquad \bar{g} = \frac{1-w}{1+w}, \qquad \sigma = -\frac{1 + 3w}{6(1+w)}\,.
\]

Since~\eqref{Fuchsian symmetry eqs} is a Fuchsian system of
equations for the variables $\Omega^{\mathrm{g}}_{\,ab}$,
$\Omega_{\bar{\mathrm{g}}}^{\,a b}$, $\Omega^\sigma_{\,ab}$,
the solution $\Omega^{\mathrm{g}}_{\,ab} = t^gg_{[ab]}$,
\mbox{$\Omega_{\bar{\mathrm{g}}}^{\,ab} = t^{\bar{g}}
g^{[ab]}$}, $\Omega^\sigma_{\,ab} = t^\sigma \sigma_{[ab]}$ we
have in our hands, is the unique solution of~\eqref{Fuchsian
symmetry eqs} that vanishes as $t\rightarrow 0$. However, the
trivial solution is an explicit solution of~\eqref{Fuchsian
symmetry eqs} with the obvious property that it vanishes in the
limit $t\rightarrow 0$. The uniqueness result thus ensures that
$\Omega^{\mathrm{g}}_{\,ab} = t^gg_{[ab]}  \equiv 0$,
$\Omega_{\bar{\mathrm{g}}}^{\,a b} =  t^{\bar{g}} g^{[ab]}
\equiv 0$, and $\Omega^\sigma_{\,ab} = t^\sigma
\sigma_{[ab]}\equiv 0$. We thus obtain symmetry of the metric
and the extrinsic curvature, which are thus solution of the
proper Einstein-Euler evolution equations.

\begin{remark}
The proof of the symmetry of the constructed solution that is
given here is inspired by the considerations
of~\cite{Andersson&Rendall} for the stiff fluid case, i.e., $w
=1$. However, the relevant arguments
in~\cite{Andersson&Rendall} are incorrect, because the
equations in~\cite{Andersson&Rendall}, which are a variant
of~\eqref{symmetry eqs}, lack a term that is quadratic in the
extrinsic curvature. This problem can apparently be fixed,
see~\cite{Damour/Henneaux/Rendall/Weaver:2002}; we note,
however, that the arguments we follow here for $w >1$ fail for
$w = 1$. The conditions~\eqref{condisw} lead to a contradiction
for $w=1$, because in that case we find
\[
\bar{g} - \sigma > -\big(1  - \frac{8}{3(1+w)}\big) = \frac{1}{3}\:,\qquad
\bar{g} - \sigma < 1 - \frac{4}{3(1+w)} = \frac{1}{3} \:.
\]
In addition, the first term on the r.h.s.\ of~\eqref{bterms}
and~\eqref{aterms} behaves like $t^{(w-1)/(1+w)}$; this
function vanishes for $t \rightarrow 0$ in the case of an
ultrastiff fluid, i.e., $w >1$, but does not vanish when $w =
1$.
\end{remark}

\subsection{The constraints}

To conclude the proof of Theorem~\ref{thm2} we consider the
Einstein-Euler constraints~\eqref{constr}. We need to show that
the constructed solutions \mbox{$\mathcal{W} = \big(g_{a b},
\tr k, \tensor{\sigma}{^a_b}, \mu,u_a\big)$} of the
Einstein-Euler evolution equations satisfy the Einstein-Euler
constraints~\eqref{constr}. If this is the case, then the
solutions $\mathcal{W}$ we have obtained are solutions of the
Einstein-Euler equations~\eqref{evol}--\eqref{euler} and
Theorem~\ref{thm2} is proved.

Consider $\mathcal{W} = \big(g_{a b}, \tr k,
\tensor{\sigma}{^a_b}, \mu,u_a\big)$ and let
\begin{equation}
\Phi = -\tensor{\sigma}{^a_b} \tensor{\sigma}{^b_a} +
\textfrac{2}{3} (\tr k)^2 + R - 2 \rho \:,
\qquad
\Psi_a = -\nabla_b \tensor{\sigma}{^b_a} - \textfrac{2}{3}
\nabla_a \tr k - j_a \:,
\end{equation}
where $\rho$ and $j_a$ are given by~\eqref{rhojs}. The
Einstein-Euler evolution equations imply equations for these
constraint quantities. We have
\begin{equation}\label{constrpropa}
\partial_t \Phi = 2 \tr k \; \Phi + \nabla^a \Psi_a \:,\qquad
\partial_t \Psi_a = \tr k\; \Psi_a - \textfrac{1}{2}\,\nabla_a \Phi\:.
\end{equation}
Let $\varphi$ and $\psi$ be constants and define $\Omega^\Phi =
t^\varphi \Phi$ and $\Omega^\Psi_{\,a} = t^\psi \Psi_a$.
From~\eqref{constrpropa} we obtain the system
\begin{subequations}\label{confuchs}
\begin{align}
t \partial_t \Omega^\Phi + \big( \textfrac{4}{1 +w} - \varphi \big) \Omega^\Phi & =
\textfrac{4}{1+w} \,\big[ 1 + \textfrac{1+w}{2}\,t \: \tr k \big] \Omega^\Phi +
t^{1 + \varphi-\psi} \nabla^a \Omega^\Psi_{\,a}\:,\\[1ex]
t \partial_t \Omega^\Psi_{\,a} + \big( \textfrac{2}{1 +w} - \psi \big) \Omega^\Psi_{\,a} & =
\textfrac{2}{1+w} \,\big[ 1 + \textfrac{1+w}{2}\,t \: \tr k \big] \Omega^\Psi_{\,a}  -
\textfrac{1}{2} t^{1 + \psi- \varphi} \nabla_a \Omega^\Phi\:,
\end{align}
\end{subequations}
where $\nabla^a \Omega^\Psi_{\,a}$ can be expressed as
\begin{equation}\label{nabom}
\nabla^a \Omega^\Psi_{\,a} = (\partial_a g^{a b}) \Omega^\Psi_{\,b} + g^{a b} \partial_a \Omega^\Psi_{\,b}
+ \textfrac{1}{2} g^{a d} g^{b c} (\partial_a g_{b c}) \Omega^\Psi_{\,d} \:.
\end{equation}

It is straightforward to show that the system~\eqref{confuchs}
is Fuchsian when the constants $\varphi$ and $\psi$ satisfy a
number of conditions. (These conditions follow
from~\eqref{confuchs} in the usual manner; in~\eqref{nabom} we
use the constructed behavior of the metric and its spatial
derivatives, see~\eqref{result2}.) In particular, there exist
appropriate choices of $\varphi$ and $\psi$ that are compatible
with the additional conditions that $\Omega^\Phi= t^\varphi
\Phi$ and $\Omega^\Psi_{\,a} = t^\psi \Psi_a$ vanish in the
limit $t\rightarrow 0$. To obtain these conditions we note that
$-\!\^0\tensor{\sigma}{^a_b} \^0\tensor{\sigma}{^b_a} +
\textfrac{2}{3} (\tr \^0k)^2 - 2 \^0\mu = 0$,
cf.~\eqref{constraint equations}, which results in
\begin{align*}
\Omega^\Phi & = t^\varphi\, \big(-\tensor{\sigma}{^a_b}
\tensor{\sigma}{^b_a} + \textfrac{2}{3} (\tr k)^2 + R - 2 \rho\big) \\[1ex]
& = t^\varphi \:\big( {-2}\, t^{1-4/3(1+w) -\varepsilon/2}
\^0\tensor{\sigma}{^a_b} \tensor{\varsigma}{^b_a} +
\textfrac{4}{3}\, t^{2-10/3(1+w) - \varepsilon} (\tr \^0k) \,\kappa\,+\\[1ex]
& \qquad\qquad \qquad+ \^0R - 2 \,t^{1-10/3(1+w) -\varepsilon} \nu -2 (1+w) \^0\mu \^0u^2
+ \text{higher order in $t$}\,\big),
\end{align*}
where ``higher order in $t$'' denotes terms that converge to
zero faster than at least one of the terms that are given
explicitly. From~\eqref{asybehgenw} we see that it is the
curvature term that determines the leading behavior of
$\Omega^\Phi$ in $t$. We obtain $\Omega^\Phi \in
O(t^{\varphi-4/3(1+w)})$. Analogously, we have
\begin{align*}
\Omega^\Psi_{\,a} & = t^\psi\, \big({-\nabla_b} \sigma^b{}_a -
\textfrac{2}{3}\nabla_a \tr k - j_a\big) \\[1ex]
& =
 t^\psi \:\big( -(\nabla_b-\^0\nabla_b) \^0\tensor{\sigma}{^b_a}
- t^{1-4/3(1+w)-\varepsilon/2} \^0\nabla_b \tensor{\varsigma}{^b_a}
- \textfrac{2}{3} t^{2-10/3(1+w)-\varepsilon} \nabla_a \kappa \\[1ex]
& \qquad\quad - (1+w) \^0\mu \,(\sqrt{1+ u^2} - 1 ) \^0u_a
- (1+w) t^{4-10/3(1+w) - 2\varepsilon} \^0\mu
\sqrt{1+u^2}\; v_a \\[1ex]
& \qquad\quad -(1+w) t^{1-10/3(1+w) -\varepsilon} \nu \^0u_a  +
\text{higher order in $t$} \:\big)\,.
\end{align*}
The leading order comes from the term $\^0\nabla_b
\tensor{\varsigma}{^b_a}$. We thus obtain $\Omega^\Psi_{\,a}
\in O(t^{\psi + (3w-1)/3(1+w) -\varepsilon/2})$. The required
conditions are satisfied if the constants $\varphi$ and $\psi$
obey the inequalities
\begin{subequations}\label{constr ineq}
\begin{align}
\frac{4}{3(1+w)} < \varphi < \frac{4}{1+w} = \frac{12}{3(1+w)} \:, & \qquad
-\big(1 - \frac{4}{3(1+w)} - \frac{\varepsilon}{2} \big) < \psi < \frac{2}{1+w} = \frac{6}{3(1+w)} \:,
\\
1 + \varphi-\psi - \frac{4}{3(1+w)}> 0 \:, & \qquad  1 + \psi- \varphi > 0 \:.
\end{align}
\end{subequations}

Assuming that $\varepsilon$ is small enough (which we can
always do) we are able to ensure that the required conditions
are satisfied; an explicit choice of constants is $\varphi =
\psi = 5/3(1+w)$. Then~\eqref{confuchs} is a Fuchsian system of
equations for the variables $\Omega^\Phi$, $\Omega^\Psi_{\,a}$,
and $\Omega^\Phi = t^\varphi \Phi$ and $\Omega^\Psi_{\,a} =
t^\psi \Psi_a$ represent a solution of that system that
vanishes as $t\rightarrow 0$. However, since the trivial
solution is an explicit solution of~\eqref{confuchs}, the
uniqueness result guarantees that $\Omega^\Phi= t^\varphi \Phi
\equiv 0$ and $\Omega^\Psi_{\,a} = t^\psi \Psi_a \equiv 0$,
i.e., the constraints $\Phi \equiv 0$ and $\Psi_a \equiv 0$ of
the Einstein-Euler system are identically satisfied.

The solution $\mathcal{W} = \big(g_{a b}, \tr k,
\tensor{\sigma}{^a_b}, \mu,u_a\big)$ of~\eqref{result2} is thus
a solution of the Einstein-Euler
equations~\eqref{evol}--\eqref{euler} and Theorem~\ref{thm2} is
proved.


\end{document}